\documentclass[preprint,showpacs,preprintnumbers,amsmath,amssymb]{revtex4}
\usepackage{amsmath}
\usepackage{bm}

\begin{document}

\preprint{BA-03-07}

\title{Explaining Why the $u$ and $d$ Quark Masses are Similar}

\author{S.M. Barr}
\email{smbarr@bxclu.bartol.udel.edu}
\author{I. Dorsner}
\email{dorsner@physics.udel.edu}
\affiliation{
Bartol Research Institute\\
University of Delaware\\
Newark, DE 19716}

\begin{abstract}
An approach is suggested for modeling quark and lepton masses and mixing
in the context of grand unified theories that explains the curious fact
that $m_u \sim m_d$ even though $m_t \gg m_b$. The structure of the quark
mass matrices is such as to allow a non-Peccei-Quinn solution of the
Strong CP Problem. \end{abstract}


\maketitle

\newpage

It is well known that the grand unification group $SU(5)$ relates the
mass matrices of the down quarks and charged leptons. There is some
empirical support for the existence of such a relationship in the
fact that when the fermion masses are extrapolated to the GUT scale in
the MSSM one finds $m_b \cong m_{\tau}$ \cite{Chanowitz:1977ye,begd},
$m_s \cong m_{\mu}/3$, and $m_d \cong 3 m_e$ \cite{gj}. However, the
pattern of masses of the up quarks
is very different. One difference is that the $t$ mass is much greater
than the $b$ and $\tau$ masses, which is usually explained by saying that
the ratio of VEVs $v_u/v_d \equiv \tan \beta$ is large compared to one. Another
difference is that the interfamily mass hierarchies are much stronger
for the up quarks than for the down quarks and charged leptons
(e.g. $m_c/m_t \ll m_s/m_b$ and $m_u/m_c \ll m_d/m_s$). It is tempting
to say that the up quark mass matrix ($M_U$) is more distantly related to
the down quark and charged lepton mass matrices ($M_D$ and $M_L$) than
the latter are to each other. On the other hand, there is the tantalizing
fact that $m_u/m_d \sim 1$.

In this paper we suggest a somewhat new approach which
qualitatively explains why $m_t \gg m_b$ but $m_u \sim m_d$.
The idea is that there are ``underlying" mass matrices (denoted by the
superscript zero) whose structure is controlled by $SO(10)$ and which
satisfy $M_U^0 \sim M_D^0 \sim M_L^0 \sim M_N^0$ (it is assumed
$v_u/v_d \sim 1$),
but that a strong mixing of the third family with
vectorlike fermions at the GUT scale distorts these underlying mass
matrices in such a way that $m_b$ and $m_{\tau}$ are highly
suppressed relative
to $m_t$. This distortion does not affect the first family much, so
the masses $m_u$, $m_d$, and $m_e$ remain of the same order.

The approach we will describe has several other virtues: (a) It can
be realized in models with very few parameters. (b) It dovetails with
the ideas of Ref.~\cite{nb} for solving the Strong CP Problem. And (c) it
implements the ``lopsided" mass matrix approach to explaining large
neutrino mixing angles \cite{bb96,lopsided}.

To understand the idea, consider an $SU(5)$ model (which will later be 
assumed to descend
from $SO(10)$) that has three families of fermions
in ${\bf 10}_i + \overline{{\bf 5}}_i + {\bf 1}_i$, with mass terms
of the form $(M^0_U)_{ij} {\bf 10}_i {\bf 10}_j
+ (M_D^0)_{ij} {\bf 10}_i \overline{{\bf 5}}_j
+ (M_L^0)_{ij} \overline{{\bf 5}}_i {\bf 10}_j
+ (M_N^0)_{ij} \overline{{\bf 5}}_i {\bf 1}_j$.
The matrices $M_D^0$ and $M_L^0$ come from the VEV of a $\overline{{\bf 5}}$
of Higgs, and $M_U^0$ and $M_N^0$ (the neutrino Dirac mass matrix) come from
the VEV of a ${\bf 5}$ of Higgs. Suppose that there are also for each family a
vectorlike pair of quark/lepton multiplets, denoted $\overline{{\bf 5}}'_i$
and ${\bf 5}'_i$ and having superheavy mass terms
$A_{ij} \overline{{\bf 5}}'_i {\bf 5}'_j + B_{ij} \overline{{\bf 5}}_i
{\bf 5}'_j$. ($A_{ij} \sim B_{ij} \sim M_{GUT}$.) There is then mixing between
the ordinary three families and the vectorlike fermions, more specifically
the mixing is between the $\overline{{\bf 5}}_i$ and the
$\overline{{\bf 5}}'_i$. (A similar idea, but with mixing among fermions in
${\bf 10}$'s of $SU(5)$ was used in \cite{bb96}. However, the models
proposed there were very different in character from the present models.)
With the mass terms specified above, we may write
\begin{equation}
\label{matrix}
\left( \overline{{\bf 5}} \;\; \overline{{\bf 5}}' \right) {\cal M}_F
\left( \begin{array}{c} {\bf 10} \\ {\bf 5}' \end{array}
\right) = \left( \overline{{\bf 5}}_i \overline{{\bf 5}}'_i \right)
\left( \begin{array}{cc} (M^0_F)_{ij}  & B_{ij} \\ 0 & A_{ij} \end{array}
\right) \left( \begin{array}{c} {\bf 10}_j \\ {\bf 5}'_j \end{array}
\right),
\end{equation}
\noindent
where $F = L$ or $D$, and $M_F^0$ is either $M_L^0$ or $(M_D^0)^T$,
depending on whether the fermions in $\overline{{\bf 5}}$ are
$\ell^-_L$ or $d^c_L$.

In order to find
the light fermion mass matrices in the effective low-energy theory,
we must do a unitary transformation ${\cal M}_F = {\cal U} {\cal M}_F^0$
that eliminates the off-diagonal block $B$ in
the full mass matrix given in Eq.~(\ref{matrix}). Such a transformation is
\begin{equation}
{\cal U} = \left( \begin{array}{cc} \Lambda & - \Lambda x \\
x^{\dag} \overline{\Lambda} & \overline{\Lambda} \end{array} \right),
\end{equation}
\noindent
where $x \equiv B A^{-1}$, $\Lambda \equiv (I + x x^{\dag})^{-1/2}$, and
$\overline{\Lambda} = (I + x^{\dag} x)^{-1/2}$.
(To check the unitarity of ${\cal U}$ it is useful
to note that $x^{\dag} \Lambda = \overline{\Lambda}
x^{\dag}$ and $x \overline{\Lambda} = \Lambda x$.) This gives the
result for the low energy mass matrices
\begin{subequations}
\label{ml:and:md}
\begin{eqnarray}
M_L & = & \Lambda M_L^0, \\
M_D & = & M_D^0 \Lambda^T.
\end{eqnarray}
\end{subequations}
\noindent
Basically, the hermitian matrix $\Lambda$ describes the mixing of
$\overline{{\bf 5}}_i$
with $\overline{{\bf 5}}'_i$. It appears on the left in the equation for
$M_L$ since $(M_L)_{ij}$ couples to $\overline{{\bf 5}}_i {\bf 10}_j$.
It appears on the right in the equation for $M_D$ since $(M_D)_{ij}$
couples to ${\bf 10}_i \overline{{\bf 5}}_j$. For the Dirac neutrino masses
we have
\begin{equation}
\left( \overline{{\bf 5}} \;\; \overline{{\bf 5}}' \right) {\cal M}_N
\left( \begin{array}{c} {\bf 1} \\ {\bf 5}' \end{array}
\right) = \left( \overline{{\bf 5}}_i \overline{{\bf 5}}'_i \right)
\left( \begin{array}{cc} (M^0_N)_{ij}  & B_{ij} \\ 0 & A_{ij} \end{array}
\right) \left( \begin{array}{c} {\bf 1}_j \\ {\bf 5}'_j \end{array}
\right),
\end{equation}
\noindent
giving
\begin{equation}
\label{mn}
M_N  = \Lambda M_N^0.
\end{equation}
\noindent
Since the masses of the up-type quarks come from a
${\bf 10}_i {\bf 10}_j$ coupling of the fermions, they are not affected
by the mixing of the $\overline{{\bf 5}}_i$ with the
$\overline{{\bf 5}}'_i$. Consequently,
\begin{equation}
\label{mu}
M_U  = M_U^0.
\end{equation}

Before we discuss how the structure we have described can help us
explain the magnitudes of quark and lepton masses and mixings, we note
that it is exactly the kind of structure
that is used in the solution of the Strong CP Problem proposed
in \cite{nb}. The idea there was the following. Suppose that CP is a
symmetry of the lagrangian that is spontaneously broken, and that the
VEV that breaks CP appears in the off-diagonal matrix $B$ in Eq.~(\ref{matrix}),
but not elsewhere in the quark mass matrices. Then $M_D^0$ and $A$
are real, and it is easily shown that the determinant of the full mass
matrix ${\cal M}_D$ is therefore real. Also real, of course, is
the determinant of $M_U$. Thus, at tree level,
the phase $\overline{\theta}$ is zero. At higher order, these matrices can
receive complex corrections that induce a non-vanishing $\overline{\theta}$,
but these may be made small. (In SUSY, there can be contributions to
the $\overline{\theta}$ parameter that are harder to make small, for
example, one-loop corrections to the gluino mass \cite{dlk}. How large these
are depends upon how SUSY is broken. These contributions are not
a problem in theories with gauge-mediated SUSY breaking, for example.
We imagine that whatever mechanism resolves the usual SUSY flavor and
SUSY CP problems will also suppress these extra contributions to 
$\overline{\theta}$.)
On the other hand, since $B$ is a complex matrix,
so is the matrix $x = B A^{-1}$ and the matrix $\Lambda =
(I + x x^{\dag})^{-1/2}$. Consequently, the mass matrix of the light
three families of down-type quarks in the effective low-energy theory,
given by $M_D = M_D^0 \Lambda^T$, is also complex, which means that in
general there is a non-vanishing Kobayashi-Maskawa phase.

In short, the structure in Eq.~(\ref{matrix}) allows a spontaneously generated phase
in the matrix $B$ to contribute to $\delta_{KM}$ but not at tree level to
$\overline{\theta}$. This can also enhance the predictivity of
models by reducing the number of parameters, since one can assume that
all parameters in
$M_L^0$, $M_D^0$, $M_U^0$, $M_N^0$, and the right-handed Majorana matrix
$M_R$ are real, and that the only
phase (and only one is needed) comes from $\Lambda$. This is the assumption
we shall make in the illustrative model we present below.

Returning to the issue of mass and mixing hierarchies, let us assume that
the matrix $\Lambda$ that characterizes the mixing of $\overline{{\bf 5}}_i$
with $\overline{{\bf 5}}'_i$, has the form
\begin{equation}
\label{lambda}
\Lambda \cong \left( \begin{array}{ccc} 1 & 0 & 0 \\
0 & 1 & \lambda t \\ 0 & \lambda t^* & \lambda \end{array} \right),
\end{equation}
\noindent
where the real parameter $\lambda \ll 1$ and the complex parameter
$t$ has magnitude of order one. As we shall see shortly, it is the smallness
of $\lambda$ that gives rise to $m_b, m_{\tau} \ll m_t$, while the
$|t| \sim 1$ explains the large atmospheric neutrino mixing.
The phase of $t$, the only phase in the model, is what produces the KM phase.
We shall see later that the form in Eq.~(\ref{lambda}) is easy to obtain.

To illustrate our basic approach we now present a toy model
in which the underlying mass matrices have the following simple ``textures":
\begin{equation}
\label{mu0:md0:mn0:ml0}
\begin{array}{cccccc}
M_U^0 & = & \left( \begin{array}{ccc} 0 & \delta & \delta' \\
\delta & \epsilon_u & 0 \\ \delta' & 0 & 1 \end{array} \right) m_U, \;\;\;\;
& M_D^0 & = & \left( \begin{array}{ccc} 0 & \delta & 0 \\
\delta & \epsilon_d & 0 \\ 0 & 0 & 1 \end{array} \right) m_D, \\
& & & & & \\
M_N^0 & = & \left( \begin{array}{ccc} 0 & \delta & \delta' \\
\delta & 3 \epsilon_u & 0 \\ \delta' & 0 & 1 \end{array} \right) m_U, \;\;\;\;
& M_L^0 & = & \left( \begin{array}{ccc} 0 & \delta & 0 \\
\delta & 3 \epsilon_d & 0 \\ 0 & 0 & 1 \end{array} \right) m_D,
\end{array}
\end{equation}
\noindent
where $\delta, \delta' \ll \epsilon_u, \epsilon_d \ll 1$.
The similarity of these four matrices is assumed to come from $SO(10)$.
In $SO(10)$ one would have the ${\bf 10}_i + \overline{{\bf 5}}_i + {\bf 1}_i$
come from a ${\bf 16}_i$, whereas the extra vectorlike fermions
$\overline{{\bf 5}}'_i + {\bf 5}'_i$ could come from a ${\bf 10}_i$.

The textures in Eq.~(\ref{mu0:md0:mn0:ml0}) can be obtained from simple $SO(10)$ operators.
In particular, we assume that the 33 elements come from a term of the form
$h_{33} {\bf 16}_3 {\bf 16}_3 \langle {\bf 10}_H \rangle$. Thus, what we
have called $m_U$ and $m_D$ in Eq.~(\ref{mu0:md0:mn0:ml0}) are given by $m_U = h_{33} \langle
H_u({\bf 10}) \rangle$, and $m_D = h_{33} \langle
H_d({\bf 10}) \rangle$. If the two Higgs doublets of the MSSM came purely
from the ${\bf 10}_H$, i.e. if $H_u = H_u({\bf 10})$ and $H_d =
H_d({\bf 10})$, then we would have $m_U/m_D = \tan \beta$. However, one
expects in a realistic $SO(10)$ model that $H_u$ and $H_d$ will come from
a mixture of several $SO(10)$ Higgs multiplets. Thus, we may write
$H_d = \cos \gamma_d \; H_d({\bf 10}) + \sin \gamma_d H_d(other)$ and
$H_u = \cos \gamma_u \; H_u({\bf 10}) + \sin \gamma_u H_u(other)$.
Inverting these, we obtain $m_D =  h_{33} v \cos \gamma_d \cos \beta$
and $m_U =  h_{33} v \cos \gamma_u  \sin \beta$. Therefore, the usual
$\tan \beta$ parameter of the MSSM is given by $\tan \beta =
(m_U/m_D)(\cos \gamma_d/\cos \gamma_u)$.
From Eq.~(\ref{mu0:md0:mn0:ml0}) one sees that the top quark mass is $m_t \cong m_U$.
Therefore, $1/\cos \gamma_u \cong h_{33} (v/m_t) \sin \beta$, and we may
write $\tan \beta = (m_U/m_D) [ \cos \gamma_d (v/m_t) \sin \beta] h_{33}$.
The expression in the square brackets is less than or equal to 1,
and $h_{33}$, which is a Yukawa coupling in the $SO(10)$ theory, cannot
be much larger than 1 without destroying the perturbativity of the
theory below the Planck scale. Thus, the value of the parameter $m_U/m_D$,
which can be determined by fitting the quark and lepton masses, puts an
upper bound on $\tan \beta$. We shall find that $m_U/m_D \cong 2$, so with
$h_{33} = 1.5$ to 2, the value of $\tan \beta$ is consistent with the
experimental lower limit of 3 \cite{tanbeta}

Now, given Eqs.~(\ref{ml:and:md}), (\ref{lambda}), and (\ref{mu0:md0:mn0:ml0}), one has
\begin{equation}
\label{md}
M_D = \left( \begin{array}{ccc} 0 & \delta & \delta \lambda t^* \\
\delta & \epsilon_d & \epsilon_d \lambda t^* \\
0 & \lambda t & \lambda \end{array} \right) m_D,
\end{equation}
\noindent
and
\begin{equation}
\label{ml}
M_L = \left( \begin{array}{ccc} 0 & \delta & 0 \\
\delta & 3 \epsilon_d & \lambda t \\ \delta \lambda t^* & 3 \epsilon_d \lambda t^* &
\lambda \end{array} \right) m_D.
\end{equation}
\noindent
Of course, from Eq.~(\ref{mu}) one sees that $M_U$ is already given in Eq.~(\ref{mu0:md0:mn0:ml0}).

Simply by inspecting these matrices one can observe several significant
facts. First, the masses of $m_b$ and $m_{\tau}$ are suppressed by
the small parameter $\lambda$, whereas $m_t$ is not, so that
$m_b, m_{\tau} \ll m_t$ can be explained without requiring that
$m_U/m_D$ be extremely large. Second, the masses of the first family
will be almost unaffected by the parameter $\lambda$, so that $m_d$ and
$m_e$ will not be similarly suppressed compared to $m_u$. Indeed
for $m_U/m_D$ of order one, $m_u \sim m_d$, as
observed. Third, there emerges naturally the ``lopsided" structure discussed
in many recent papers \cite{lopsided}. That is, we see that the 23 element
of $M_L$
is much larger than its 32 element, whereas for $M_D$ the opposite is the
case. This comes directly from the fact that $M_D = M_D^0 \Lambda^T$ whereas
$M_L = \Lambda M_L^0$. This lopsided structure explains why the atmospheric
neutrino mixing angle (which gets a contribution from $(M_L)_{23}/(M_L)_{33}$)
is of order $|t| \sim 1$, whereas the corresponding quark mixing $V_{cb}$
(which gets a contribution from $(M_D)_{23}/(M_D)_{33}$) is only of
order $\epsilon_d |t| \ll 1$.

One can read off from the simple forms in Eqs.~(\ref{md}) and (\ref{ml}) the following
approximate relations that hold at the GUT scale:
\begin{equation}
\label{masses}
\begin{array}{lll}
m_t \cong m_U, \;\; & m_c \cong \epsilon_u \; m_U, \;\; & m_u \cong
(\delta^2/\epsilon_u) \; m_U, \\
& & \\
m_b \cong \lambda \sqrt{1 + |t|^2} \; m_D, \;\; & m_s \cong
(\epsilon_d/\sqrt{1 + |t|^2}) \; m_D, \;\; & m_d \cong
(\delta^2/\epsilon_d) \; m_D, \\
& & \\
m_{\tau} \cong \lambda \sqrt{1 + |t|^2} \; m_D, \;\; & m_{\mu} \cong
(3 \epsilon_d/\sqrt{1 + |t|^2}) \; m_D, \;\; & m_e \cong
(\delta^2/3 \epsilon_d) \; m_D, \\
& &
\end{array}
\end{equation}
\noindent
and
\begin{equation}
\label{mixings}
\begin{array}{ccl}
V_{cb} & \cong & (\epsilon_d/\lambda) \left( \frac{t}{1+ |t|^2} \right)
\cong (m_s/m_b) \; t, \\ & & \\
V_{us} & \cong & (\delta/\epsilon_d) - (\delta/\epsilon_u)
\cong  \sqrt{m_d/m_s} (1 + |t|^2)^{-1/4} \pm \sqrt{m_u/m_c}, \\ & & \\
V_{ub} & \cong &  - \delta' + \left(
\frac{\delta}{\lambda} \frac{t}{1 + |t|^2} \right) \left( 1 -
\frac{\epsilon_d}{\epsilon_u} \right) \cong - \delta' + V_{us} V_{cb}. \\
& & \end{array}
\end{equation}
\noindent
From the form of the $M_L$ (Eq.~(\ref{ml})) and $M_N$ (Eqs.~(\ref{mn}) and (\ref{mu0:md0:mn0:ml0}))
it can be seen that the tangent of the
atmospheric neutrino mixing angle is controlled by $t$, which therefore must
be of order one. That in turn implies, through
the equation for $V_{cb}$, that $V_{cb} \sim m_s/m_b$. This succesful
qualitative relation between the atmospheric neutrino angle, $V_{cb}$ and
$m_s/m_b$ is characteristic of lopsided models.

In this model there are eight parameters
($m_U/m_D$, $\epsilon_u$, $\epsilon_d$, $\delta$, $\delta'$, $\lambda$,
$|t|$, and $\theta_t$) to fit twelve quantities (eight mass ratios of
charged leptons and quarks, three CKM angles, and the KM phase). There are
therefore four quantitative predictions, which can be taken to be
$m_b \cong m_{\tau}$, $m_s \cong m_{\mu}/3$, $m_d \cong 3 m_e$, and
the value of the Cabibbo angle. In addition, the atmospheric angle is
predicted to be of order one, though it cannot be predicted more precisely
than that without knowing $M_R$.

We can easily determine the approximate values of most of the parameters
of the model from Eqs.~(\ref{masses}) and (\ref{mixings}). We take the values of
the quark and lepton masses and CKM mixings at the GUT scale to be
$m_t = 112$ GeV, $m_b = 0.96$ GeV, $m_{\tau} = 1.16$ GeV,
$m_c = 0.27$ GeV, $m_s = 0.015$ GeV, $m_{\mu} = 0.069$ GeV,
$m_u = 0.57$ MeV, $m_d = 0.86$ MeV, $m_e = 0.334$ MeV,
$|V_{cb}| = 0.0357$, and $|V_{us}| = 0.222$.
These are found by extrapolating experimentally determined
central values at low scale \cite{pdg} to the GUT scale using the following
procedure. First, we propagate the masses of light quarks and leptons
from $2$ GeV scale to $M_Z$ scale
using the 3-loop QCD and 1-loop QED renormalization group equations (RGEs).
Then, we perform additional running
from $M_Z$ to $m_t$ scale using the Standard Model RGEs. (The relevant
renormalization-group $\beta$ functions are summarized in Ref.~\cite{SMRGE}.)
Finally, assuming all SUSY particle masses to be degenerate at $m_t$
we run the masses and mixings to the GUT scale $M_{GUT} \approx 2 \times 10^{16}$ GeV
using the 2-loop MSSM $\beta$ functions summarized in Ref.~\cite{MSSMRGE}.
In the final running we set $\tan \beta = 3$.

The equation for $V_{cb}$ tells us
immediately that $|t| \cong V_{cb} m_b/m_s \approx 2$. From Eq.~(\ref{masses}) one
has that $(m_u m_c)/(m_d m_s) \cong (m_U/m_D)^2 \sqrt{1 + |t|^2}$,
which implies that $m_U/m_D \approx 2$. The equation $\lambda
\cong (m_b/m_t) (m_U/m_D)/\sqrt{1 + |t|^2}$ then gives $\lambda
\approx 10^{-2}$.

The value of $|\epsilon_u|$ is given approximately
by $\sqrt{m_c/m_t} \cong 2.4 \times 10^{-3}$. The equation
$\epsilon_d \cong \frac{1}{3} (m_{\mu}/m_{\tau}) \lambda (1 + |t|^2)$
gives $|\epsilon_d| \approx 10^{-3}$. It is gratifying that
$\epsilon_u$ and $\epsilon_d$ come out to be of the same order.
If we choose the relative sign of $\epsilon_u$ and $\epsilon_d$ to be
negative, then we get a good fit to the Cabibbo angle:
$V_{us} \cong \sqrt{m_d/m_s} (1 + |t|^2)^{-1/4} + \sqrt{m_u/m_c}
\cong (0.2)(0.7) + (0.05) = 0.2$.

The value of $\delta$ is determined from $\delta^2 \cong m_u m_c/m_t^2$
to be $10^{-4}$. Finally, the parameter $\delta'$ and the
phase of $t$ can be determined from the real and imaginary parts of
$V_{ub}$. Specifically, one has
\begin{equation}
V_{ub}/(V_{us} V_{cb}) = 1 - \delta' e^{- i \theta_t}/|V_{us} V_{cb}|.
\end{equation}

One gets a fairly reasonable fit from the following
values of the parameters of the model: $m_U/m_D = 2.03$,
$\lambda = 1.03 \times 10^{-2}$,
$|t| = 1.45$, $\epsilon_U = 2.38 \times 10^{-3}$, $\epsilon_D = - 2.14
\times 10^{-3}$, $\delta = 1.12 \times 10^{-4}$. The resulting masses and
mixings and the experimental values extrapolated to the GUT scale are
compared in Table~\ref{tab:table1}.
\begin{table}[h]
\caption{\label{tab:table1}The values of the quark and charged lepton
masses and the CKM angles $V_{cb}$ and $V_{us}$ at the GUT scale
in the model (with parameter values given in text), compared to the
experimental values extrapolated to the GUT scale. Extrapolation is done
taking all SUSY particles to be degenerate at $m_t$ and assuming $\tan \beta = 3$.
Masses are given in units of GeV.}
\begin{ruledtabular}
\begin{tabular}{lcc}
&{\rm model}&{\rm experiment}\\
\hline
$m_u$ & $0.000587$ & $0.000570$ \\
$m_c$ & $0.268$ & $0.269$ \\
$m_t$ & $112$ & $112$ \\
$m_d$ & $0.00092$ & $0.00086$ \\
$m_s$ & $0.0238$ & $0.0150$ \\
$m_b$ & $0.998$ & $0.956$ \\
$m_e$ & $0.000318$ & $0.000334$ \\
$m_{\mu}$ & $0.0684$ & $0.0690$ \\
$m_{\tau}$ & $1.00$ & $1.16$ \\
$|V_{us}|$ & $0.19$ & $0.22$ \\
$|V_{cb}|$ & $0.032$ & $0.036$ \\
\end{tabular}
\end{ruledtabular}
\end{table}
It is apparent that the fit is not completely satisfactory. In particular,
the mass of the $\tau$ comes out about 15\% too small. This is typical of
grand unified theories. Simple GUTs generally predict $m_b = m_{\tau}$ at
the GUT scale, whereas the data tend to give $m_{\tau}$ about 15 to 20\%
larger than $m_b$. There are a number of ways of improving the agreement,
including supposing that $m_b$ gets corrections from sparticle loops.
Also off considerably here is $m_s$. The Georgi-Jarlskog relation
$m_s \cong m_{\tau}/3$ is built into a choice of Clebsch in this toy
model. But that relation is known to give a value of $m_s$ that is
somewhat large compared to the values favored by recent lattice calculations 
\cite{slattice}.

While this model does not give a perfect fit, it is simple enough to
illustrate the basic idea we are proposing in a transparent way.
It seems likely that models based on these ideas can be found that
give better fits. Another possibility that might be realized
in this approach is a ``doubly lopsided" model \cite{bb02}.
One could imagine, for
example, that the matrix $\Lambda$ had the form
\begin{equation}
\Lambda \cong \left( \begin{array}{ccc}
1 & 0 & \lambda t_1 \\ 0 & 1 & \lambda t_2 \\ \lambda t_1^* &
\lambda t_2^* & \lambda \end{array} \right),
\end{equation}
\noindent
with $\lambda \ll 1$ and $|t_i| \sim 1$.
If the underlying matrix $M_L^0$ has a hierarchical form, with the
33 element being the largest, then the effective low-energy mass matrix
$M_L = \Lambda M_L^0$ would have the doubly lopsided form, with the
13, 23, and 33 elements all being of the same order. This is known to be
able to give in a simple way the correct ``bi-large" pattern of neutrino
mixing angles, with $U_{e3}$ being small \cite{bb02}.

We now turn to the question of whether the form of $\Lambda$ given in
Eq.~(\ref{lambda}) can arise naturally. Consider the special case where
$B$ is diagonal and where the only non-zero elements of $A$
are the diagonal elements and the 23 element: $B_{ij} = b_i \delta_{ij}$
$A_{ij} = a_i \delta_{ij} + a_4 \delta_{i2} \delta_{j3}$. Then it is easily
found that for $b_1 < a_1$, $b_2 < a_2$, and $b_3 > a_3$, the matrix
$\Lambda$ has the form given in Eq.~(\ref{lambda}) with
$\lambda \cong |a_3|^2/(|a_3|^2 + |b_3|^2)$, and $t = -(b_3^* b_2 a_4)/
(a_2 |a_3|^2)$. Of course, there are other forms of $A$ and $B$ that also
give Eq.~(\ref{lambda}). Another simple example is that $A$ is diagonal and $B$ has
nonzero diagonal elements and 23 element.

In conclusion, we have found a framework that differs from most ``texture"
models of quark and lepton masses in several respects. First, it can
partially explain the fact, usually treated as an accident, that
$m_u \sim m_d, m_e$, while also giving $m_t \gg m_b, m_{\mu}$. This it
does, not by requiring $\tan \beta$ to be large, which might be 
somewhat unnatural, but by mixing the $b$ and $\tau$ strongly
with vectorlike fermions at the GUT scale. Second, it combines
predictive textures with a structure that realizes a non-axion
solution to the strong CP problem proposed many years ago \cite{nb}. 
By allowing most of the parameters to be real, even though CP is violated,
it has the potential of giving very predictive models. And it
gives rise naturally to the ``lopsided" kind of structure that has
been proposed to explain the largeness of $U_{\mu 3}$ relative to
$V_{cb}$ \cite{lopsided}. 

The toy model we have described illustrates the essential ideas in a
transparent way. However, it would be good to find a model which is more
predictive and which does a better job fitting certain quantities, especially
$m_s$. It would also be interesting to investigate further models of
this type that are ``doubly lopsided" \cite{bb96, bb02}.

\end{document}